\documentstyle[floats,aps]{revtex}

\begin{document}
\draft

\catcode`\@=11 \catcode`\@=12 
\twocolumn[\hsize\textwidth\columnwidth\hsize\csname@twocolumnfalse\endcsname
\title{From Composite Fermions to Calogero-Sutherland Model: Edge of Fractional 
Quantum Hall Liquid and the Dimension Reduction}
\author{Yue Yu }
\address{Institute of Theoretical Physics, Chinese Academy of  Sciences, 
P. O. Box 2735,
Beijing 100080, China}

\maketitle
\begin{abstract}
 
We derive a microscopic model describing the low-lying edge excitations in the fractional quantum Hall liquid with $\nu=\frac{\nu^*}{\tilde\phi\nu^*+1}$. For $\nu^*>0$, it is found that the composite fermion model reduces to an SU$(\nu^*)$ Calogero-Sutherland model in a dimension reduction, whereas it is not exact soluble for $\nu^*<0$. However, the ground states in both cases can be found and the low-lying excitations can be shown the chiral Luttinger liquid behaviors.  On the other hand,
 we shows that the finite temperature behavior of $G-T$ curve will deviate from the prediction of the chiral Luttinger liquid. We also point out that the suppression of the `spin' degrees of freedom agrees with very recent experiments by Chang et al. The two-boson model of Lee and Wen is described microscopically.

\end{abstract}

]

The basic characteristic of the quantum Hall states is the incompressibility
of the two-dimensional (2-d) electron system in a strong perpendicular
magnetic field \cite{Laugh}. While there is a finite energy gap for
particle-hole excitations in the bulk, the low-lying gapless excitations are
located at the edge of the quantum Hall liquid \cite{Halp}. A phenomenological
picture of the edge states of fractional quantum Hall (FQH) effect is beyond
the Fermi liquid framework and known as the chiral Luttinger liquid (CLL) 
\cite{Wen,Kane}. Recently, the edge excitations of FQH effect (FQHE) were
studied by several groups numerically \cite{Rez} or in accordance with the
Calogero-Sutherland model (CSM) \cite{ECS} as well as the composite fermion
(CF) picture in the Hartree approximation \cite{CFP}.

In previous works \cite{Yu1}, we have given a microscopic model of the CF
for the edge excitations at the filling factor $\nu=1/m ,~m={\rm odd~integer}
$. It was seen that at $\nu=1/m$ the CF system \cite{Jain} reduces to the
original CSM \cite{CS} with a dimension reduction. And the low-lying
excitations of the edge, then, are proven to be governed by a CLL. We further applied
this microscopic theory to analyze the tunneling experiment \cite{Chang} and
got a better fit to the features of the measured tunneling conductivity versus temperature curve \cite{Zheng}. However, the CLL is applied not only to the one channel edge excitation, i.e. , at $\nu=1/m$ but also to the multi-channel case, for
example, at $\nu=\frac{\nu^*}{\tilde\phi\nu^*+1}$ for integer $\nu^*$ and
even number $\tilde\phi>0$ \cite{Wen}. Furthermore, a systematic
experimental study of the current-voltage ($I-V$) characteristic for the
electron tunneling between a metal and the edge of a 2-d electron gas at a
fractional filling factor $\nu$ shows a continuous non-Ohmic exponents $
\alpha=1/\nu$, i.e., $I\sim V^\alpha$ \cite{Gray}. This is contradict with
the prediction of the CLL theory in which, say, $I\sim V^3 $ for the primary
filling factor being 1/3. Several authors have made their efforts in
explaining these phenomena \cite{expl,Lee}. Lee and Wen recently proposed a
two-boson model for FQHE regime to explain the experimental result in which the spin velocity is much slower
than the charge's and then the long time behavior shows the exponent $%
\alpha=1/\nu$ while the short time behavior complies with the Fermi
statistics of the electrons \cite{Lee}.

In this article, we propose a generalization of our previous work in the
following sense. On the one hand, we consider a 2-d anyon model with SU$(\nu
^{*})$ `spin' symmetry and to see how it will be under a dimension reduction
to one. On the other hand, this dimension reduction leads to a microscopic
derivation of the CLL at $\nu =\frac{|\nu ^{*}|}{\tilde{\phi}|\nu ^{*}|\pm 1}
$. It is seen that the edge theory at $\nu =\frac{\nu ^{*}}{\tilde{\phi}\nu ^{*}+1}$ is basically described by the SU$(\nu ^*)$ CSM, i.e., the $\nu ^{*}$-channel theory
whose low-lying excitations are separated into charge and `spin' channels where
the $\nu ^{*}-1$ spins run in the same direction as the charge's. However, it is not the case for that at $\nu =\frac{|\nu ^{*}|}{
\tilde{\phi}|\nu ^{*}|-1}$ ($\nu ^{*}<0$). The edge theory at these filling
factors does not correspond to an exact soluble model. Fortunately, we could
still find a good approximation ground state wave function and then the $
|\nu ^{*}|$ channel low-lying excitations where the charge channel runs in
the opposite direction to the spin channels. We show that the exclusion
statistics between the channels is described by the $K$ matrices \cite{WenZ}%
. Thus, using the bosonization procedure developed in \cite{Wu} the system
with the exclusion statistics matrix $K$ can be taken as the fixed point of
the multi-channel Luttinger liquid. From the radial wave equation of the
system, one can show that the residual magnetic field provides a gap between
right- and left-moving modes in each channel of these low-lying
excitations. This verifies the CLL at the $T\to 0$ limit. From this
microscopic picture, the spin and charge velocities can be estimated and one
finds that $v_s^{*}<<v_\rho ^{*}$. This supports the two-boson theory of Lee
and Wen. Very recently, Chang et al observed a plateau behavior in $\nu_{\rm edge}$ approximately between $2/5$ and $2/7$ around $1/3$ \cite{chang1}, which seems to support Lee-Wen's theory. We also argue that our model at $\nu^*\to \infty$
may apply to describe the edge state at $\nu=1/2$. 

The CLL is one side in which the ground state behavior of the
system is well described. On the other hand, the finite temperature behavior
of the system shows a crossover from the CLL to the Fermi liquid \cite{Chang}%
. We have explained this crossover at $\nu=1/m$ \cite{Zheng}. Here, we
predict that this crossover could also appear for other filling fractions
like $\nu =2/5$ if applying our argument for $\nu=1/m$ to the charge
channel of the multi-channel system.

To have a microscopic picture of the edge states in FQHE, we begin with the
CF theory of FQHE \cite{Jain}. The CF description of the bulk states has
been greatly investigated \cite{HLR} and been focused by many recent studies 
\cite{RCF}.  However, to understand the gapless excitations of the FQH system,
it might be not important to know the details of the bulk behavior because
all bulk excitations are gapped. After integrating over the all gapped
states, the low energy behavior of the system is governed by the edge CFs.
For a disc geometry, the edge CFs are restricted in a circular strip near
the boundary with its width $\delta R(\vec{r})\ll R$, the radius of the
disc. The edge Hamiltonian of the CFs in the unit of $\hbar =e/c=2m^{*}_e=1$
reads \cite{Yu1} 
\begin{eqnarray}
H_e &=&\sum_i\biggl[-\frac{\partial ^2}{\partial r_i^2}+(-\frac i{r_i}\frac %
\partial {\partial \varphi _i}-\frac{B^{*}}2r_i)^2  \label{HE} \\
&+&\frac{\tilde{\phi}^2}{4R^2}\sum_i(\sum_{j\not{=}i}\cot \frac{\varphi _{ij}%
}2)^2  \nonumber \\
&-&\frac{\tilde{\phi}}R\sum_{i<j}\cot \frac{\varphi _{ij}}2\cdot i(\frac %
\partial {\partial r_i}-\frac \partial {\partial r_j})-\frac 1R\frac \partial
{\partial r_i}\biggr]  \nonumber \\
&+&\sum_{i<j}V(\vec r_i-\vec r_j)+\sum_i U_{eff}(\vec r_i)+O(\delta R/R). 
\nonumber
\end{eqnarray}
where the external potential $U_{eff}$ is the effective potential including
the interaction between the edge and bulk particles. $m^*_e$ is the edge CF effective
mass. Here we have used the polar coordinate $x_i=r_i\cos \varphi
_i,~y_i=r_i\sin\varphi _i$ and bulk mean field approximation has been
imposed. The residual magnetic field $B^{*}=\frac \nu {\nu ^{*}}B$. The
manifestation of CF picture is that the $\nu ^{*}$ denotes the highest
Landau level index of CF in the residual magnetic field. Our central focus
is to solve the many-body problem $H_e\Psi _e(z_1\sigma _1,...,z_{N_e}\sigma
_{N_e})=E\Psi _e(z_1\sigma _1,...,z_{N_e}\sigma _{N_e})$ where $\sigma
_i=1,...,\nu ^{*}$ is the Landau level index which we call spin hereafter.
And the many-body wave function $\Psi _e$ has to be consistent with the bulk
state. In the previous works \cite{Yu1}, we have presented an example to
solve the problem at $\nu =1/m$, i.e., $\nu ^{*}=1$, and see that the edge
ground states can be directly related to Laughlin's wave function in the
bulk. However, it is not applied to a general FQH state with $\nu =\frac{\nu
^{*}}{\nu ^{*}\tilde{\phi}+1}$.

We start from $\nu ^{*}>0$. To the zero order of $V$, we first switch off
this interaction. Without loss of the generality, one takes the trial wave
function is of the form 
\begin{eqnarray}
&&\Psi _e(z_1\sigma _1,...,z_{N_e}\sigma _{N_e})  \nonumber \\
&=&\exp \biggl[\frac i2\sum_{i<j}t_{\sigma _i\sigma _j}\frac{r_i-r_j}R\cot 
\frac{\varphi _{ij}}2-\frac A2\frac{r_i-r_j}R(J_{\sigma _i}-J_{\sigma _j})%
\biggr]  \nonumber \\
&&\times f(r_1,...,r_{N_e})\Psi _s(\varphi _1\sigma _1,...,\varphi
_{N_e}\sigma _{N_e}),  \label{TRI}
\end{eqnarray}
where $t_{\sigma _i\sigma _j}$ is a parameter matrix to be determined to match the particle statistics after the dimension reduction and so
is $A$. $J_\sigma $ is the spin quantum number. The radial wave function $f$
is symmetric and the azimuthal wave function $\Psi _s$ is anti-symmetric in
the particle exchange. To be consistent with the bulk wave function, the
azimuthal wave function takes its form, 
\begin{eqnarray}
&&\Psi _s(\varphi _1\sigma _1,...,\varphi _{N_e}\sigma
_{N_e})=\prod_{i>j}\phi _{ij}\cdot \prod_k\xi _k^{J_{k_\sigma }},  \nonumber
\\
&&\phi _{ij}=|\xi _i-\xi _j|^{\tilde{\phi}}(\xi _i-\xi _j)^{\delta _{{\sigma
_i}{\sigma _j}}}\exp \{i\frac \pi 2{\rm sgn}(\sigma _i-\sigma _j)\},
\label{AWF}
\end{eqnarray}
where $\xi _i=e^{i\varphi _i}$. In the one-dimensional (1-d) limit, taking $%
\delta R/R\to 0$ in (\ref{HE}) after acting on $\Psi _e$, the Hamiltonian on 
$\Psi _e$ yields $H_{cs}$ on $\Psi _s$ with 
\begin{equation}
H_{cs}=\sum_i(i\frac \partial {\partial x_i}+\frac{B^{*}}2R)^2+\frac{\pi ^2}{%
L^2}\sum_{i<j}\frac{\tilde{\phi}(\tilde{\phi}+P_{\sigma _i\sigma _j})}{[\sin
(\frac{\pi x_{ij}}L)]^2},  \label{HCS}
\end{equation}
where $x_{ij}=x_i-x_j$, $\varphi _i=\frac{2\pi x_i}L$ and $L=2\pi R$ is the
size of the boundary. $P_{\sigma _i\sigma _j}$ is the spin exchange operator 
\cite{HH}. To arrive at (\ref{HCS}), the matrix $t_{\sigma _i\sigma _j}$ is
taken as $\frac 12\delta _{\sigma _i\sigma _j}-\frac 14$ and $A\equiv 1/N_e$%
. The Hamiltonian (\ref{HCS}) is just the SU$(\nu ^{*})$ CSM Hamiltonian
with a constant shift to the momentum operator \cite{SUCS,HH} and the ground
state wave function is given by taking CFs in each channel have the same
number $M$ and $J_1=...=J_{\nu ^{*}}=(M-1)/2$ in $\Psi _s$. So, we see that
a 2-d anyon model with SU$(\nu ^{*})$ spin symmetry is reduced to the exact
soluble SU$(\nu ^{*})$ CSM in 1-d. Moreover, from the zeros of the
ground state wave function, one can read out the exclusion statistics matrix 
$K_{\sigma \rho }=\tilde{\phi}+\delta _{\sigma \rho }.$
Furthermore, the asymptotic Bethe ansatz (ABA)
equation which determines the psudomomentum $n_{i\sigma _i}$.
The psudomomentum $n_{i\sigma _i}$ relates to $J_{\sigma _i}$
in a complicated way and we do not show it explicitly. Because we have used
an SU$(\nu ^{*})$ symmetric form to construct the azimuthal wave function,
the ABA equations are symmetric for the spin indices. In a work characterizing the Luttinger liquid in
terms of the ideal excluson gas \cite{Wu}, we have bosonized the single
component CSM and arrived at the single channel Luttinger liquid. This
procedure can be generalized to bosonize the SU$(\nu ^{*})$ CSM. The
generalization is somewhat trivial but tedious. We do not present the
details here because the result is just as expected: a $\nu ^{*}$-channel
Luttinger liquid \cite{HH,Wen}.

The azimuthal dynamics has two directions: the right- and
left-moving excitations. However, the radial equation of the wave function
ranges the right-moving sector out of the lower-lying excitation sector.
To see the chirality, one considers the radial wave equations. As we have done,
for the single-branch problem, the radial problem can be reduced
to the single-particle one except the $
n_{i\sigma_i} $ (shorted as $n$) is related by the ABA equations. As the
first step, we consider an ideal case in which the effective potential takes
an infinite wall at the boundary of the 2-d system. The eigenenergy of the
single-particle wave function $g_+(y)=g(r_n-R)$ with $n>0$ is $|n|\hbar\omega^*_c$
lower than that of $g_-(y)=g(r_n-R)$ with $n<0$\cite{Yu1}, where $r_n=\sqrt{%
\frac{2|n|}{B^*}}$. Therefore, the states with the negative $n$ are ranged
out of the low-lying state sector. This implies that only the left-moving
mode in the azimuthal dynamics belongs to the low-lying excitation sector.
The width of the wave function $g_+$ is several times of $R_c^*$, the
cyclotron radius of the CF in the effective field. For the highest spin $%
\nu^*$, one can see that the eigenenergy of $g_+(0)$ ($r_{n_{\nu^*}} =R$) is 
$\varepsilon_{R\nu^*}=(2(\nu^*-1)+\frac{3}{2}) \hbar\omega^*_c$ since the
wave function vanishes at $r=R$ while $\varepsilon_{n_{\nu^*}}=((\nu^*-1)+ 
\frac{1}{2})\hbar\omega^*_c$ for $R-r_n\gg R_c^*$, a harmonic oscillator
energy and coinciding with the mean field theory applied to the bulk states.
The gapless excitations appear when $|R-r_n|\sim R_c^*$ . Using a
perturbative calculation, one has the bare excitation energy is $%
\varepsilon_0=\varepsilon_{n_{\nu^*}}-\varepsilon_{R\nu^*} \sim v_F\frac{eB^*%
}{c}(r_n-R)$ with the Fermi velocity $v_F=\frac{\pi\hbar\rho_0}{m^*}(\tilde%
\phi+\nu^*)$ where $m^*$ is the bulk CF effective mass.

The SU$(\nu ^{*})$ symmetry forces all other channels of the edge
excitations have the same velocity as the $\nu ^{*}$-th channel. This is
contradict to the recent edge tunneling experiment \cite{Gray}. Lee and Wen 
\cite{Lee} argued that this inconsistency could be dispelled if the spin
mode velocities are much smaller than the charge's. ( Charge mode is given
by $\rho _c\propto \sum \rho _\sigma $ and so on.) In our model, there are
two factors to change the sound velocity $v_c$ which equals to $v_F$ before
considering those factors. First, the confining potential is
smooth in real samples. The sound velocity is given by $\tilde
v_c=\frac{d\hbar\omega_k}{dk}$ with $\hbar\omega_k=(\nu^*+1/2)\hbar\omega^*_c
+U_{eff}(kl_B^{*2})$.  $U_{eff}(r)$ is slowly varying in the real sample,
i.e., $ \partial_rU_{eff}(r)\ll\hbar\omega_c^*/l_B^*$ 
For the simplicity, we assume the edge potential has 
the form that $U_{eff}=0$ in the deep of the sample
and $U_{eff}=\infty$ outside of the sample. Near the boundary, it can be expanded
in $r-R'$, $U_{eff}=eE(r-R')+O((r-R')^2)$ where the induced electric field $E$
describes the edge sharpness. The exact value of $R'$ is not important for
$R-R'=\delta R$ . Thus, due to $\partial_rU_{eff}(r)
\approx eE\ll \hbar\omega_c^*/l_B^*$ \cite{comm} 
and $U_{eff}(kl^{*2}_B)\approx eE(kl^{*2}_B-R')$, one has
\begin{equation}
v_c\approx eEl^{*2}_B/\hbar=cE/B^*\ll k_F/m^*\sim v_F,\label{VL}
\end{equation}  
i.e., the real sound velocity ${v}_c$ is much smaller than $v_F$. 
The other factor to affect the sound velocity is the interaction $V$ which violates the SU$(\nu^{*})$ symmetry. The interaction is of the form $\rho _\sigma V_{\sigma
\sigma ^{\prime }}\rho _{\sigma ^{\prime }}$ for $\rho_\sigma$ being the 
CF density operators. It can be
rewritten as $\rho _cV_\rho \rho _c+\rho _sV_{sc}\rho _c+\rho
_sV_{ss^{\prime }}\rho _{s^{\prime }}$ where $\rho _c$ is the charge density
wave and $\rho _s$ are the $\nu ^{*}-1$ channels of spin density wave. All
the interactions stem from the electron-electron interaction, which yields $%
V_{ss^{\prime }},V_{sc}\ll V_\rho $. We can assume $V_{ss^{\prime }}=V_{sc}=0$%
. Therefore, only the charge density wave velocity is renormalized by the
interaction $V_\rho $. Finally, we have the renormalized velocities 
\begin{equation}
v_\rho ^{*}={v}_c+V_\rho ,~~~~v_s^{*}={v}_c.\label{vv}
\end{equation}
We see that the smooth edge potential suppresses the spin wave velocity
whereas the interaction $V_\rho $ is not changed. Here we consider the
interaction to be short range. It is reasonable to take $V_\rho $ in the order of the average Coulomb interaction, $\frac{2\pi e^2l_B}{\epsilon\hbar}$, which is 
proportional to the inverse of the bulk CF effective mass and of the order of 
$v_F$. According to (\ref{VL}) and (\ref{vv}), $v_\rho ^{*}\approx V_\rho \gg v_s^{*}$
which is just what Lee and Wen predicted \cite{Lee}. In an experimental observation to be published \cite{chang1}, Chang et al observed a plateau behavior in the vicinity of $\nu_{\rm edge}=1/3$ in the CLL exponents versus $\nu_{\rm edge}$. However, this plateau suddenly disappears if either $\nu_{\rm edge}>\frac{2}{2\times 2+1}=2/5$ or $<\frac{2}{4\times 2-1}=2/7$ approximately. That is to say, before the spin degrees of freedom are turned on, the system is controlled by the single channel CLL for $\nu=1/m$ while the spin breaks the plateau behavior.  Therefore, this observation
is consistent with the picture of the spin velocity suppression.

We have discussed the ground state behavior. Chang et al have shown a
temperature crossover in the tunneling conductivity $G$ versus temperature
in $\nu=1/3$ \cite{Chang}. This crossover from the CLL to Fermi liquid can
be explained by using our microscopic theory in which the dressed energy
either is smooth if $T\gg k_F$ or is of a refraction at Fermi surface if $
T\ll k_F$ \cite{Zheng}. This form of the dressed energy still holds for the
charge channel of the multi-channel edge state at $\nu=\frac{\nu^*}{\tilde
\phi\nu^*+1}$. Therefore, it is expected to be the general feature that the measurement in $G(T,V)$  obeys the Ohmic law for $T\gg V$ while the CLL ground state emerges for $T\ll V$. Then a crossover from the CLL to Fermi liquid for can be observed as temperature varying from $T\ll V$ to $T\gg V$. 

Turn to much subtle problem at $\nu^*<0$, i.e., $\nu=\frac{|\nu^*|}{|\nu^*|%
\tilde\phi-1}$ . To solve the problem, we make an anyon transformation for
the edge CF with the statistical parameter $b\delta_{\sigma\sigma^{\prime}}$
where b is a real number which is given by a solution of equation $(\tilde%
\phi-1)b^2+2\tilde\phi+2=0$. Although the 1-d limit model is still not
soluble after this transformation, we may attract the low-lying excitation
sector by using a trial wave function. A trial wave function enlightened by
the bulk wave function may be taken of the form 
\begin{eqnarray}
&&\Psi_e(z_1\sigma_1,...,z_{N_e}\sigma_{N_e})  \nonumber \\
&&=\exp\biggl[\frac{i}{2} \sum_{i<j}s_{\sigma_i\sigma_j}\frac{r_i-r_j}{R}%
\cot \frac{\varphi _{ij}}{2}-\frac{A}{2}\frac{r_i-r_j}{R} (J_{\sigma_i}-J_{%
\sigma_j})\biggr]  \nonumber \\
&&\times f(r_1,...,r_{N_e})
\Psi_s(\varphi_1\sigma_1,...,\varphi_{N_e}\sigma_{N_e}),  \label{TRI1}
\end{eqnarray}
where $s_{\sigma_i\sigma_j}=-\frac{1}{4}+c\delta_{\sigma_i\sigma_j}$ with $%
c=(\frac{1}{4\tilde\phi}-\frac{1}{2})b-\frac{1}{2}$. The azimuthal wave
function $\Psi_s$ can be set according to the $K$ matrix of the bulk state
and reads 
\begin{eqnarray}
&&\Psi_s(\varphi_1\sigma_1,...,\varphi_{N_e}\sigma_{N_e})
=\prod_{i>j}\phi_{ij}\cdot\prod_k \xi_k^{J_{k_\sigma}}  \nonumber \\
&&\phi_{ij}=|\xi_i-\xi_j|^{\tilde\phi}(\xi_i-\xi_j)^{-\delta_{{\sigma_i} {%
\sigma_j}}}  \nonumber \\
&&\times\exp\{i\frac{\pi}{2}[{\rm sgn}(\sigma_i-\sigma_j) +b{\rm sgn}%
(i-j)\delta_{\sigma_i\sigma_j}]\}.  \label{AWF1}
\end{eqnarray}
This wave function is indeed an eigen wave function of the anyon-transformed
Hamiltonian if $\delta R/R\to 0$. Taking a suitable set of the quantum
numbers as that for $\nu^*>0$, we have the ground state wave function. The
important matter is that the exclusion statistics of the azimuthal wave
function is given by the expected bulk $K$ matrix $K_{\sigma\sigma^{\prime}}=%
\tilde\phi-\delta_{\sigma\sigma^{\prime}}$. According to this exclusion
statistics matrix and our bosonization approach, we can finally arrive the
CLL theory in which the charge mode travels in the opposite direction than
the $\nu^*-1$ spin-modes, which, in the clean edge , leads to the absence of
the edge equilibration . However, the effective potential $U_{eff}$ includes
all possible external potential. Of them, a most relevant one is the random
impurity potential. Kane et al have shown that this random potential drives
the edge to a stable fixed point and restores the edge equilibration \cite
{Kane}.

Finally, we discuss the possible application of the model presented here to
the edge state at $\nu=1/2$. 
Because there are the gapless bulk excitations, 
it is not clear how to identify the edge states. Our formalism provides 
a way to define the edge state at $\nu=1/2$ 
if considering the SU($\nu^*$) model at $\nu^*\to \infty$. 
As discussed before, the spin velocity is also
suppressed, which is consistent with the power-law $I-V$ behavior in 
the experimental observation.
There are two problems to be clarified. On the one hand, the bulk excitations
can not be integrated over from the lower-lying excitation sector. However, they might
not affect the $I-V$ behavior because the buck CF has a very large (divergent) effective
mass at $\nu=1/2$. On the other hand, it is not clear whether the edge excitation
at $\nu=1/2$ is chiral and our model does not support the chirality because of $B^*=0$
in this case.

In conclusion, we have proposed a microscopic model of edge excitations for
FQHE at $\nu =\frac{\nu ^{*}}{\nu ^{*}\tilde{\phi}+1}$ which is realized
through a dimension reduction from a 2-d anyon model to the SU$(\nu ^{*})$
CSM for $\nu ^{*}>0$ while there is no exact soluble counterpart for $\nu
^{*}<0$. The low-lying excitations for both $\nu ^{*}>0$ and $\nu ^{*}<0$,
however, are proven to be described by the CLL. We argued that the two-boson
theory of Lee-Wen is valid to explain the experiments by Grayson et al in
FQH regime and consistent with the plateau behavior of the CLL exponents recently
observed. For the finite temperature, we predicted a crossover from CLL to Fermi liquid for $\nu ^{*}>0$ as observed in the one-channel case.

The author thanks T. Xiang and G. M. Zhang for useful discussions. This work
was supported in part by the NSF of China. He is grateful to A. M. Chang for sending
Ref. \cite{chang1}.


\vspace{-0.2in}

\begin{references}

\vspace{-0.4in}

\bibitem{Laugh}  R. B. Laughlin, Phys. Rev. Lett. {\bf 50}, 1395 (1983).

\bibitem{Halp}  B. I. Halperin, Phys. Rev. B {\bf 25}, 2185 (1982).

\bibitem{Wen}  X. G. Wen, Phys. Rev. Lett. {\bf 64}, 2206 (1990); Phys. Rev.
B {\bf 41}, 12838 (1990).

\bibitem{Kane}  C. L. Kane, M. P. A. Fisher and J. Polchinski, Phys. Rev.
Lett. {\bf 72}, 4129 (1994). C. L. Kane and M. P. A. Fisher, Phys. Rev. B 
{\bf 51}, 13449 (1995).

\bibitem{Rez}  E. H. Rezayi and F. D. M. Haldane, Phys. Rev. B {\bf 50},
6924 (1994). A. Cappelli, C. Mendez, J. M. Simonin, G. R. Zemba,
cond-mat/9806238, to appear in Phys. Rev. B (1999)

\bibitem{ECS}  F. D. M. Haldane, Bull. Am. Phys. Soc. {\bf 37}, 164(1992);
Phys. Rev. Lett. {\bf 67}, 937 (1991); S. Mitra and A. H. MacDonald, Phys.
Rev. B {\bf 48}, 2005 (1993); P. J. Forrester and B. Jancovici, J. Phys.
(Pairs) {\bf 45}, L583 (1994); N. Kawakami, Phys. Rev. Lett. {\bf 71},
275(1993); A. D. Veigy and S. Ouvry. Phys. Rev. Lett. {\bf 72}, 121(1994).
H. Azuma and S. Iso, Phys. Lett. B{\bf 331}, 107 (1994). S. Iso, D.  Karabali and B. Sakita
Nucl. Phys. B{\bf 388 },  700 (1992) .
\bibitem{CFP}  L. Brey, Phys. Rev. B {\bf 50}, 11861 (1994); D. B.
Chkolvskii, Phys. Rev. B {\bf 51}, 9895 (1995).

\bibitem{Yu1}  Y. Yu and Z. Y. Zhu, Commun. Theor. Phys. {\bf 29}, 351
(1998). Y. Yu, W. J. Zheng and Z. Y. Zhu, Phys. Rev. B {\bf 56}, 13279
(1997).

\bibitem{Jain}  J. K. Jain, Phys. Rev. B {\bf 41}, 7653 (1990) and
references therein.

\bibitem{CS}  F. Calogero, J. Math. Phys. {\bf 10} 2197 (1967). B.
Sutherland, J. Math. Phys. {\bf 12}, 246, 251 (1971).

\bibitem{Chang}  A. M. Chang, . L. N. Pfeiffer and K. W. West, Phys. Rev.
Lett. {\bf 77}, 2538 (1996).

\bibitem{Zheng}  W. J. Zheng and Y. Yu, Phys. Rev. Lett. {\bf 79}, 3242
(1997).

\bibitem{Gray}  M. Grayson, D. C. Tsui, L. N. Pfeiffer, K. W. West
and A. M. Chang, Phys. Rev. Lett. {\bf 80}, 1062 (1998).

\bibitem{expl}  A. V. Shytov, L. S. Levitov and B. I. Halperin, Phys. Rev.
Lett. {\bf 80}, 141 (1998). S. Conti and G. Vignale, cond-mat/9801318. U.
Zulicke and A. H. Macdonald, cond-mat/9802019.

\bibitem{Lee}  D. H. Lee and X. G. Wen, cond-mat/9809160; K. Imura,
cond-mat/9812400.

\bibitem{WenZ}  X. G. Wen and A. Zee, Phys. Rev. B {\bf 46 }, 2290 (1992).

\bibitem{Wu}  Y. S. Wu and Y. Yu. Phys. Rev. Lett. {\bf 75}, 890 (1995).

\bibitem{chang1} A. M. Chang, M. K. Wu, C. C. Chi, L. N. Pfeiffer and K. W. West,
`Plateau behavior in the chiral Luttinger liquid exponents', to appear. 

\bibitem{HLR}  B. I. Halperin, P. A. Lee and N. Read, Phys. Rev B {\bf 47},
7312 (1993). V. Kalmeyer and S. C. Zhang, Phys. Rev. B {\bf 46}, 9889 (1992).

\bibitem{RCF}  R. Shankar and G. Murthy, Phys. Rev. Lett.{\bf 79}, 4437
(1997). V. Pasquier and F. D. M. Haldane, Nucl. Phys. B {\bf 516},
719(1998).D. H. Lee, Phys. Rev. Lett. {\bf 80}, 4745 (1998). N. Read, Phys.
Rev. B {\bf 59}, 8084 (1998); A. Stern, B. I. Halperin, F. von Oppen and S.
H. Simon, cond-mat/9812135.

\bibitem{HH}  Z. N. C. Ha and F. D. M. Haldane, Phys. Rev. B {\bf 46}, 9359
(1992).

\bibitem{SUCS}  Y. Kuramoto and H. Yokoyama, Phys. Rev. Lett. {\bf 67}, 1338
(1991); N. Kawakami, Phys. Rev. B {\bf 46}, 1005 (1992); K. Hikami and M.
Wadati, J. Phys. Soc. Jpn. {\bf 62}, 469 (1993); J. A. Minahan and A. P.
Polychronakos, Phys. Lett. B {\bf 302} 265 (1993).
\bibitem{comm} The order of the magnitude $E$ has been numerically estimated 
in a recent work (H. D. Wei and Yue Yu, Commun. Theor. Phys. {\bf 32}, 321 (1999).).
The edge states are sensitive to $\frac{eE}{\hbar\omega^*/l^*_B}\sim 0.07$. Otherwise, the edge states behaviors like with the sharp edge. 
For too small $eE$, it is obvious that $U_{eff}$ behaviors like totally flat
in $R'<r<R$. For too large $eE$, it is no difference from the sharp edge potential. 
\end{references}
\end{document}